\documentclass[a4paper,10pt,twocolumn]{article}
\usepackage[utf8]{inputenc}
\usepackage{multicol,graphicx}
\usepackage{mathptmx,amsmath,amssymb}
\usepackage{bm}
\usepackage[T1]{fontenc}
\usepackage{textcomp}
\usepackage{url}
\usepackage[backend=bibtex, style=numeric,doi=false,isbn=false,url=false,maxnames=6,citestyle=numeric-comp]{biblatex}
\fontfamily{ptm}\selectfont
\bibliography{ref.bib}

\usepackage[font=small]{caption}
\usepackage{amsmath}

\newcommand{\mysection}[1]{\vspace{0.4cm} \uppercase{#1} \vspace{0.4cm}}
\newcommand{\mysubsection}[1]{\hspace{10pt}\textit{#1:}}

\begin{document}
	
\setlength{\textfloatsep}{10pt plus 1.0pt minus 2.0pt}	
\setlength{\columnsep}{1cm}


\twocolumn[%
\begin{@twocolumnfalse}
\begin{center}
	{\fontsize{14}{18}\selectfont
		\textbf{\uppercase{Adversarial Feature Learning in Brain Interfacing: An Experimental Study on Eliminating Drowsiness Effects}}\\}
	\begin{large}
        \vspace{0.6cm}
        O.~\"{O}zdenizci$^\star$, B.~Oken$^\dagger$, T.~Memmott$^\dagger$, M.~Fried-Oken$^\dagger$, D.~Erdo\u{g}mu\c{s}$^\star$\\
        \vspace{0.6cm}
        $^\star$Northeastern University, Boston, MA, USA\\
        $^\dagger$Oregon Health \& Science University, Portland, OR, USA\\
        \vspace{0.5cm}
        Corresponding author: oozdenizci@ece.neu.edu
        \vspace{0.4cm}
    \end{large}
\end{center}	
\end{@twocolumnfalse}%
]%

ABSTRACT: Across- and within-recording variabilities in electroencephalographic (EEG) activity is a major limitation in EEG-based brain-computer interfaces (BCIs). Specifically, gradual changes in fatigue and vigilance levels during long EEG recording durations and BCI system usage bring along significant fluctuations in BCI performances even when these systems are calibrated daily. We address this in an experimental offline study from EEG-based BCI speller usage data acquired for one hour duration. As the main part of our methodological approach, we propose the concept of adversarial invariant feature learning for BCIs as a regularization approach on recently expanding EEG deep learning architectures, to learn nuisance-invariant discriminative features. We empirically demonstrate the feasibility of adversarial feature learning on eliminating drowsiness effects from event related EEG activity features, by using temporal recording block ordering as the source of drowsiness variability.

\mysection{Introduction}

Majority of the limitations for electroencephalography (EEG) based brain-computer interface (BCI) systems are caused by across- and within-recording variabilities of the neural activity. While at one end daily psychological states or longitudinal motivational factors can influence BCI performance ${[1]}$, on the other hand, gradual changes in fatigue and drowsiness levels during long neural activity recordings are also known to have significant influence on the recorded EEG and hence BCI performances ${[2,3,4,5]}$. Such vigilance state fluctuations over long durations of BCI usage can eventually hinder the neural intent inference and decision making pipelines which are usually calibrated daily. Hence, they further require updating or re-calibration of the system during long standing usage. In order to address this infeasibility in brain interface system designs, to the contrary of the significant amount of work exploring neural indicators of drowsiness levels ${[6,7,8]}$, we highlight an approach towards drowsiness-invariant EEG feature extraction.

In this context, deep neural networks offer significant promise as recently popularized in the form of generic EEG feature extractors ${[9,10,11,12,13]}$. Despite being mostly investigated in offline studies ${[14]}$, deep and complex architectures of such networks are usually assumed to be capable of capturing invariant neural activity to learn generalizable EEG feature representations. However going further, there exists a significant line of work on structured adversarial model training methods for invariant representation learning in many application areas of image processing, both for generative modeling to learn attribute-invariant encoded latent representations for data augmentation ${[15,16,17]}$, as well as for discriminative models to learn task-distinguishing representations which simultaneously conceal some attribute-specific information ${[18,19]}$. In the light of this work, beyond regularizing conventional EEG feature extraction pipelines, we explore whether the recent progress in EEG deep learning architectures could be potentially extended to discover the underlying nuisance (e.g., drowsiness) invariant and BCI task-discriminative neural activity with structured, adversarial feature learning approaches.

In this study we present and assess the feasibility of adversarial feature learning on eliminating drowsiness effects (i.e., drowsiness-invariance) from EEG data to operate a BCI. We use experimental data recorded from 19 healthy participants while using an event-related potential (ERP) based BCI speller, namely the RSVP Keyboard\texttrademark~${[20]}$, in offline copy-spelling mode for five consecutive calibration blocks, for a total of approximately one hour. We use temporal calibration block ordering as a discretized variable of the source of variability in the recorded neural activity as supported by concurrently collected introspective measures of boredom and sleepiness from the participants. Accordingly, we adversarially censor a conventional convolutional neural network (CNN) architecture to obtain drowsiness-invariant ERP detection models. Finally, we present our results regarding the decoding performance gains with features extracted via adversarial model training for drowsiness-invariance.

\mysection{Adversarial Feature Learning}

We propose discriminative model training under an adversarial learning framework that aims to learn nuisance-invariant features. Particularly in this application, we aim to learn an EEG feature extractor that exploits information which is ideally independent of the drowsiness levels of the participant during calibration recordings, as well as discriminative of the participant's target versus non-target visual stimuli attention (i.e., ERP vs non-ERP). The overall purpose is to use such a task-specific invariant EEG feature extractor where drowsiness levels can fluctuate arbitrarily (e.g., later courses of the experiment session), and/or be different than the drowsiness level of the participant when the calibration EEG data is being recorded for model training. Here, we use within-session EEG recording blocks as a synthetically discretized measure of drowsiness levels. Based on the application context, this nuisance variable can vary beyond drowsiness levels, to calibration session day identification numbers (IDs) or subject IDs as well. Our notation is as follows:
\begin{itemize}
    \item $\{(\bm{X}_i,l_i,b_i)\}_{i=1}^{n}$: Model training data set,
    \item $\bm{X}_i\in\mathbb{R}^{C \times T}$: Raw EEG data recorded from $C$ channels for $T$ discretized time samples at epoch $i$,
    \item $l_i \in \{0,1,\ldots,L-1\}$: Class label corresponding to the EEG epoch for the classification problem,
    \item $b_i \in \{1,2,\ldots,B\}$: Recording block identification (ID) number as a source of drowsiness variability.
\end{itemize}

A regular discriminative CNN architecture can be composed of a convolutional feature extractor (i.e., \textit{encoder}) and a subsequent fully-connected output \textit{classifier} layer with softmax units which generates an $L$ dimensional output of normalized log-probabilities across classes. Specifically, a deterministic \textit{encoder} $f = g(\bm{X};\mu_e)$ with parameters $\mu_e$ generates feature representations $f$ which are further used as input by a \textit{classifier} modeling the likelihood $p_{\mu_c}(l \vert f)$ with parameters $\mu_c$ to estimate $l$. Training of such a CNN is performed by minimizing the cross-entropy loss (i.e., maximizing the log-likelihood).

Adversarial training of this network will be performed by censoring the features $f$ to contain as less information as possible regarding the nuisance variable $b$. Specifically, an \textit{adversary} network modeling the likelihood $p_{\mu_a}(b \vert f)$ with parameters $\mu_a$ will aim to recover the recording block IDs $b$ from the learned features in parallel. However while training the \textit{adversary} in parallel to maximize the log-likelihood for $b$, the overall objective of CNN will have an additional adversarial censoring term to force the \textit{encoder} to conceal information regarding $b$ in the learned representations $f$. During this adversarial training game, the \textit{encoder} will ideally learn features that will not be able to successfully recover $b$, but will also retain sufficient discriminative information regarding $l$. The overall objective function will be as follows:
\begin{equation}
\min_{\mu_e,\mu_c} \max_{\mu_a} \mathbb{E}[-\log p_{\mu_c}(l \vert f) + \lambda \log p_{\mu_a}(b \vert f)],
\label{eq:objective}
\end{equation}
where the features $f=g(\bm{X};\mu_e)$. Note that a higher adversarial weight $\lambda>0$ indicates enforcing stronger invariance, hence trading-off with discriminative power of the features. Here, setting $\lambda=0$ corresponds to training a regular CNN model. Optimization of the \textit{encoder-classifier} and \textit{adversary} networks based on this objective were done alternatingly with stochastic gradient descent.

\mysection{Experimental Study}

\mysubsection{Participants and Data Acquisition} Nineteen healthy subjects participated in this single-session study (six male, mean age = 33.42$\pm$13.18). Before the experiments, all participants gave their informed consent after the experimental procedure was explained to them. This study was reviewed and approved by the Oregon Health \& Science University Institutional Review Board (\#15331).

During the experiments, a DSI-24 dry-electrode EEG headset (Wearable Sensing, San Diego, CA) was used for data recordings. EEG data were recorded at 20-channels placed on the scalp according to the International 10-20 system with 300 Hz sampling frequency, Butterworth bandpass filtered at 1.5 to 42 Hz. Two ear-clip electrodes were used for reference placed on both earlobes, and FPz electrode location was used as ground.

\mysubsection{Study Design} In this offline study, during the experiments, five consecutive calibration blocks were recorded while the participants were comfortably sitting on a chair placed in front of a computer screen and were using the RSVP Keyboard\texttrademark, an EEG-based BCI speller that relies on the rapid serial visual presentation (RSVP) paradigm to visually evoke event-related brain responses ${[20]}$. Stimuli presentations are performed using twenty-eight characters including the English alphabet letters, and two additional symbols for backspace and space. During calibration block recordings, participants were instructed to visually attend to the center of the screen.

Each calibration block constituted of 100 \textit{trials} and lasted 11 minutes. As a result, pure BCI system usage duration of each participant was approximately one hour. In each trial, the task was to attend to a particular \textit{target} letter randomly determined at the beginning of the trial. Prior to each trial, the target letter cue for that trial was shown on the screen for two seconds, followed by a red cross presented for one second to indicate preparation for rapid serial visual presentation. Subsequently, trials included a series of 15 successive stimuli presentations (including the target letter) at a rate of 5 Hz at the center of the screen. Followed by a blank-screen inter-trial break, the next trial began with presentation of the new target letter. For each participant, per calibration block, experimental data constituted of 1500 stimuli presentations (100 target, 1400 non-target). EEG recording signal quality was re-examined between calibration blocks.

\begin{figure}
    \centering
    \includegraphics[width=0.46\textwidth]{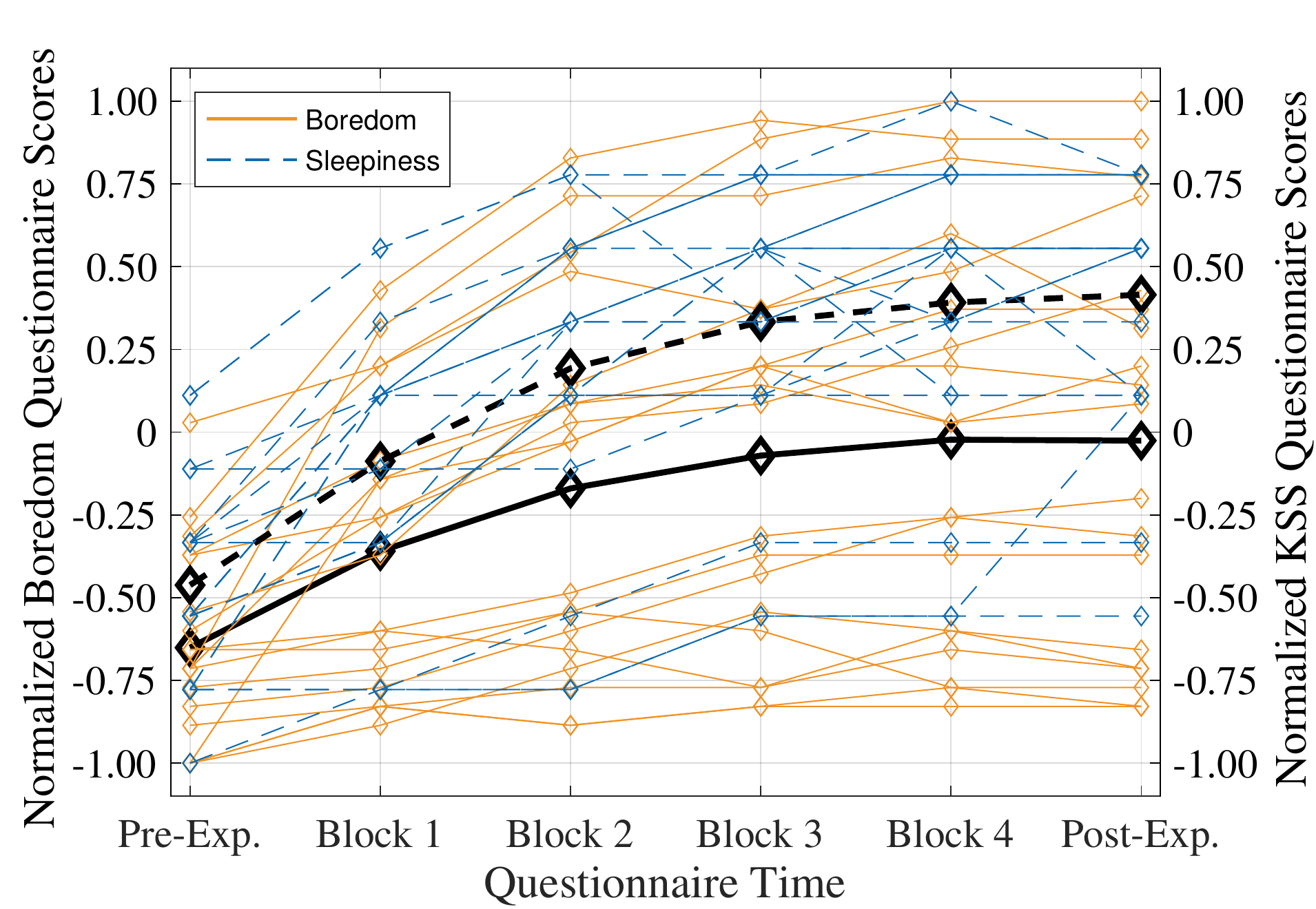}
    \caption{Normalized values of the documented introspective measures of boredom and sleepiness throughout the experiment by the 6-item Boredom Questionnaire and the Karolinska Sleepiness Scale (KSS) questionnaire. Each subjects' scores are represented by a pair of dashed blue and solid orange lines. Black and bold curves indicate the means across subjects. Greater values indicate higher boredom and sleepiness.}
    \label{fig:tests}
\end{figure}

After each calibration block, introspective measures of boredom and sleepiness are obtained based on two self-rated questionnaires: the 6-item Boredom Questionnaire ${[21]}$ and the Karolinska Sleepiness Scale (KSS) ${[22]}$. Fig.~\ref{fig:tests} presents the questionnaire results, demonstrating an overall increase in drowsiness and sleepiness levels of the participants as the experiments progressed. More experimental analyses on the neurophysiological correlates and implications of these states were previously studied based on the same experimental data set in ${[5]}$.

\mysubsection{EEG Data Analysis} EEG signal epochs were extracted from [0-600] ms post-stimuli time intervals. This resulted in EEG data matrices of dimensions $C$=20 channels times $T$=180 discretized time samples. As the only pre-processing step, all trial channels were normalized to have zero mean and scaled to [-1,1] range by dividing with the absolute maximum value ${[12]}$. No offline channel selection or artifact correction was performed.

\mysubsection{Adversarial Network Architecture} Proposed approach is applicable to any existing neural network based EEG feature extraction paradigm. For our demonstrations, the convolutional \textit{encoder} we implemented was adapted from the EEGNet architecture ${[13]}$. We modified the architecture based on the input representations of our data set (e.g., sampling rate, number of channels, etc.). We used 8 temporal convolution units and two depthwise spatial convolution units per each temporally convolved feature map (i.e., the EEGNet-8,2 convention from ${[13]}$). Convolutional \textit{encoder} specifications were as follows:
\begin{itemize} \setlength\itemsep{-0.1em}
    \item Input EEG data dimensionality: (20,180)
    \item Convolution Block 1:
    \vspace{-0.5em}
    \begin{itemize} \setlength\itemsep{-0.1em}
        \item 8 x Temporal Conv1D (1,90)
        \item Batch Normalization
        \item 2 x Spatial DepthwiseConv1D (20,1)
        \item Batch Normalization + ReLU Activation
        \item Average Pooling: (1,3)
        \item Dropout (0.25)
    \end{itemize}
    \vspace{-0.4em}
    \item Convolution Block 2:
    \vspace{-0.5em}
    \begin{itemize} \setlength\itemsep{-0.1em}
        \item 16 x Spatio-Temporal Conv1D (1,15)
        \item Batch Normalization + ReLU Activation
        \item Dropout (0.25)
        \item Flatten
    \end{itemize}
    \vspace{-0.4em}
    \item Output feature vector dimensionality: (1,240)
\end{itemize}

Afterwards, the \textit{encoder} output feature vectors $f$ are used as input both to a \textit{classifier} and an \textit{adversary} network which were simply constructed as single fully-connected layers with $L$ and $B$ softmax output units respectively.

\begin{figure*}
	\centering
	\includegraphics[trim=4cm 0cm 4cm 0cm, width=0.99\textwidth]{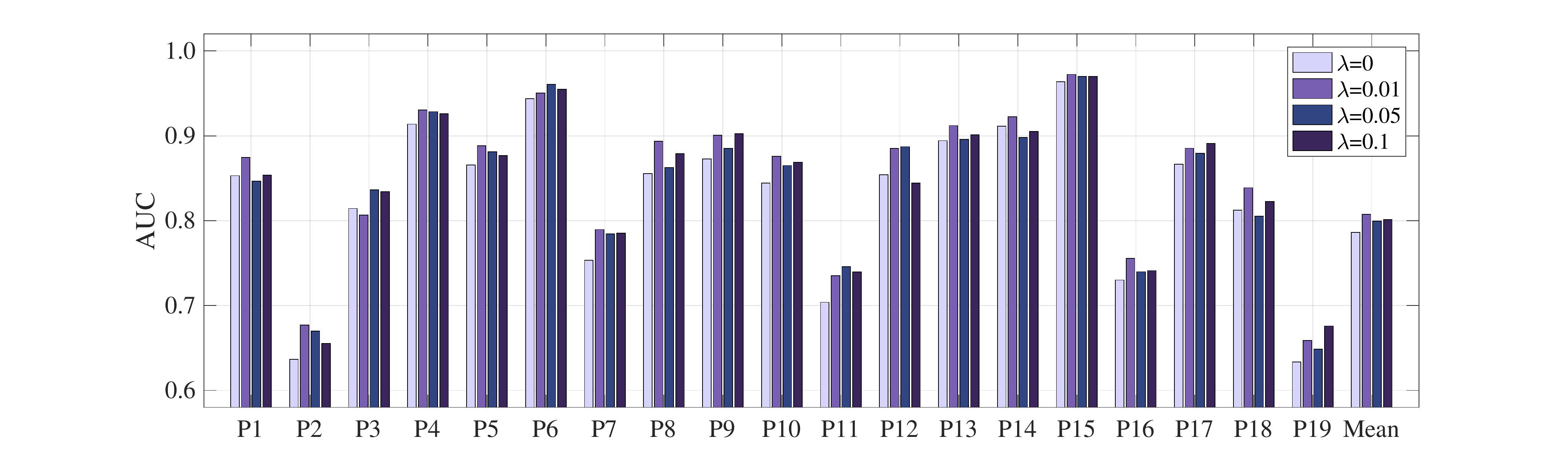}
    \caption{Area-under-the-curve (AUC) values calculated via the ROC curves of target detection on the testing set, for each subject, for varying values of $\lambda$. Regular CNN training corresponds to $\lambda=0$, whereas $\lambda>0$ indicates adversarial feature learning schemes.}
\vspace{-0.1cm}
\end{figure*}

\mysubsection{Implementation} We analyzed each participant's data individually and generated model learning and testing splits based on their calibration blocks. Specifically, the last 2 calibration blocks of the recording session was determined as the \textit{testing set}, and models were trained based on the remaining first part of the data set (i.e., first 3 calibration blocks). From that part of the data set, we split 10\% of the trials (randomly stratified by $b\in\{1,2,3\}$, $l\in\{0,1\}$ and pooled) as the \textit{validation set}, and the remaining as the \textit{training set}. This resulted in 4050 training, 450 validation and 3000 testing epochs.

At this stage we were not interested in investigating an optimal choice of adversarial regularization weight parameter $\lambda$, and we simply chose various arbitrary values ($\lambda\in\{0,0.01,0.05,0.1\}$) for our pilot demonstrations. An intuitive way to determine is to repeat the model learning step while observing validation set \textit{classifier} accuracies since higher adversarial regularization weights will start to impact discriminative performance.

All analyses were performed with the Chainer deep learning framework ${[23]}$. Networks were trained with 50 trials per batch for 200 epochs with early stopping based on validation loss, and parameter updates were performed once per batch with Adam ${[24]}$. Total number of network parameters to be estimated were 6,160. On average, total elapsed time for GPU model trainings were 57.0$\pm$20.7 seconds for $\lambda=0$ and 57.8$\pm$16.5 seconds for $\lambda>0$.

\mysection{Experimental Results}

We observe relative contributions of adversarial feature learning with respect to its non-adversarial counterpart (i.e., $\lambda=0$) in learning drowsiness-invariant features from three calibration blocks of data, to utilize the BCI speller in later phases of the experiments when drowsiness levels further increase based on introspective measurements. After model trainings were complete, three-class decoding \textit{adversary} network performances on the \textit{validation set} averaged across subjects were observed as in Tab.1. In the non-adversarial case, an \textit{adversary} is only trained in parallel (i.e., with no adversarial loss feedback) to observe the amount of exploited drowsiness-specific information (i.e., leakage). High deviations for the non-adversarial model points to the fact that deep models can indeed exploit subject- and drowsiness-variant activity in decoding (up to $93\%$ leakage with a participant). Stronger adversarial regularization can reduce block ID decoding accuracies to the $33\%$ chance level.

\vspace{0.3cm}
\begin{small}
Table 1: Three-class calibration block identification accuracies of the \textit{adversary} network after model training on the \textit{validation set} with varying $\lambda$ values, averaged across subjects.
\vspace{-0.1cm}
\renewcommand{\arraystretch}{1.4}
	\begin{center}
	\begin{tabular}{c c}
	\hline
	\textbf{Adversarial Weight} & \textbf{Validation Adversary Acc.} \\
	\hline
	$\lambda=0$ & $48.8\%\pm13\%$ \\
	$\lambda=0.01$ & $44.0\%\pm5\%$ \\
	$\lambda=0.05$ & $39.7\%\pm5\%$ \\
	$\lambda=0.1$ & $38.2\%\pm4\%$\\
	\hline
	\end{tabular}
	\end{center}
\end{small}

An ideal BCI speller classifier would allow high sensitivity with a low false alarm rate. Hence, using the complete testing set EEG epoch samples, performance was estimated based on the area-under-the-curve (AUC) values calculated through the receiver operating characteristic (ROC) curves. Calculation of the ROC curves were based on sensitivity for target class detection versus false alarm rates on target detection. Fig.~\ref{fig:tests} demonstrates an overall summary of the AUCs for each participant's data with varying feature learning schemes. The mean bars denote the average AUCs across participants as $78.6\%\pm20.7\%$ with $\lambda=0$, and $80.7\%\pm20.9\%$, $79.9\%\pm20.7\%$ and $80.1\%\pm20.7\%$ with $\lambda\in\{0.01,0.05,0.1\}$ respectively. Both on average and on an individual level, exploiting drowsiness-invariant features via the adversarial feature learning framework yields better AUCs than the $\lambda=0$ bars. These differences were also found statistically significant by non-parametric Wilcoxon signed-rank tests on the null hypothesis of zero median across paired differences between non-adversarial and adversarial model AUCs: $\lambda=0$ versus $\lambda=0.01$ ($p<0.0002$), $\lambda=0.05$ ($p<0.002$), and $\lambda=0.1$ ($p<0.0004$). This finding strongly relates to the nature of the regular deep learning frameworks where the models can arbitrarily exploit nuisance-specific, vigilance based temporal information to build the \textit{classifier}. Structured adversarial learning yields improvements, but in some cases trading-off with discriminative performance when adversarial censoring is strong. Cross-validating $\lambda$ based on the \textit{validation set} \textit{classifier} and \textit{adversary} accuracies is a feasible option.

\mysection{Discussion}

We present an adversarial invariant representation learning framework in the context of the recent progress on EEG deep learning studies. We argue that such an adversarial regularization approach could extend discriminative EEG feature extractors to further censor nuisance-specific variability in the data. In our pilot experimental studies, we assess the feasibility of adversarial feature learning particularly on eliminating drowsiness effects from event related EEG features extracted via a CNN-based encoder. Preliminary results demonstrate the feasibility of our approach in learning drowsiness-invariant representations on decoding performance.

In this work, it is important to highlight that we do not present a deep neural network model for feature learning in BCIs, but rather an adversarial regularization scheme that has not been considered in the EEG deep learning studies. We demonstrate our approach based on a recent CNN architecture (i.e., EEGNet ${[13]}$). Our approach is applicable to any existing neural network-based EEG feature extraction paradigm for BCIs. On a broader interpretation, one can explore deep invariant latent representation learning by disentangling other specific attributes (i.e., nuisance variables) from the representations via adversarial censoring such as variables based on calibration session days, or subject identifiers for across-subject transfer learning. Recently, we explored a similar approach using adversarial variational autoencoders in our preliminary work on subject-to-subject transfer feature extractor models ${[25]}$, and for across-recording EEG-based biometric identification models ${[26]}$. Future work on our approach requires investigating the neurophysiological interpretations of the learned invariant features within the networks and what it promises for BCIs.

\mysection{Acknowledgment}

Our work is supported by NSF (IIS-1149570, CNS-1544895, IIS-1715858), DHHS (90RE5017-02-01), and NIH (R01DC009834).

\mysection{references}

${[1]}$ Nijboer F, Birbaumer N, K{\"u}bler A. The influence of psychological state and motivation on brain-computer interface performance in patients with amyotrophic lateral sclerosis--a longitudinal study. Frontiers in Neuroscience. 2010;4(55).\\

${[2]}$ Ergenoglu T, et al. Alpha rhythm of the EEG modulates visual detection performance in humans. Cognitive Brain Research. 2004;20(3):376-383.\\

${[3]}$ K{\"a}thner I, Wriessnegger SC, M{\"u}ller-Putz GR, K{\"u}bler A, Halder S. Effects of mental workload and fatigue on the P300, alpha and theta band power during operation of an ERP (P300) brain-computer interface. Biological Psychology. 2014;102:118-129.\\

${[4]}$ Myrden A, Chau T. Effects of user mental state on EEG-BCI performance. Frontiers in Human Neuroscience. 2015;9:308.\\

${[5]}$ Oken B, Memmott T, Eddy B, Wiedrick J, Fried-Oken M. Vigilance state fluctuations and performance using brain computer interface for communication. Brain-Computer Interfaces. 2019 (to appear).\\

${[6]}$ Shi L-C, Lu B-L. EEG-based vigilance estimation using extreme learning machines. Neurocomputing. 2013;102:135-143.\\

${[7]}$ Martel A, D{\"a}hne S, Blankertz B. EEG predictors of covert vigilant attention. Journal of Neural Engineering. 2014;11(3):035009.\\

${[8]}$ Wei C-S, Lin Y-P, Wang Y-T, Lin C-T, Jung T-P. A subject-transfer framework for obviating inter-and intra-subject variability in EEG-based drowsiness detection. NeuroImage. 2018;174:407-419.\\

${[9]}$ Cecotti H, Graser A. Convolutional neural networks for P300 detection with application to brain-computer interfaces. IEEE Transactions on Pattern Analysis and Machine Intelligence. 2011;33(3):433-445.\\

${[10]}$ Stober S, Sternin A, Owen AM, Grahn JA. Deep feature learning for EEG recordings, in Proc. International Conference on Learning Representations, 2016.\\

${[11]}$ Bashivan P, Rish I, Yeasin M, Codella N. Learning representations from EEG with deep recurrent-convolutional neural networks, in Proc. International Conference on Learning Representations, 2016.\\

${[12]}$ Schirrmeister RT, et al. Deep learning with convolutional neural networks for EEG decoding and visualization. Human Brain Mapping. 2017;38(11):5391-5420.\\

${[13]}$ Lawhern V, et al. EEGNet: a compact convolutional neural network for EEG-based brain-computer interfaces. Journal of Neural Engineering. 2018.\\

${[14]}$ Lotte F, et al. A review of classification algorithms for EEG-based brain-computer interfaces: a 10 year update. Journal of Neural Engineering. 2018;15(3):031005.\\

${[15]}$ Edwards H, Storkey A. Censoring representations with an adversary, in Proc. International Conference on Learning Representations, 2016.\\

${[16]}$ Louizos C, Swersky K, Li Y, Welling M, Zemel R. The variational fair autoencoder, in Proc. International Conference on Learning Representations, 2016.\\

${[17]}$ Lample G, et al. Fader networks: Manipulating images by sliding attributes, in Proc. Annual Conference on Neural Information Processing Systems (NIPS), 2017, 5967-5976.\\

${[18]}$ Xie Q, Dai Z, Du Y, Hovy E, Neubig G. Controllable invariance through adversarial feature learning, in Proc. Annual Conference on Neural Information Processing Systems (NIPS), 2017, 585-596.\\

${[19]}$ Louppe G, Kagan M, Cranmer K. Learning to pivot with adversarial networks, in Proc. Annual Conference on Neural Information Processing Systems (NIPS), 2017, 981-990.\\

${[20]}$ Orhan U, Hild KE, Erdo\u{g}mu\c{s} D, Roark B, Oken B, Fried-Oken M. RSVP keyboard: An EEG based typing interface, in Proc. IEEE International Conference on Acoustics, Speech, and Signal Processing, 2012, 645-648.\\

${[21]}$ Markey A, Chin A, Vanepps EM, Loewenstein G. Identifying a reliable boredom induction. Perceptual and Motor Skills. 2014;119(1):237-253.\\

${[22]}$ Gillberg M, Kecklund G, Akerstedt T. Relations between performance and subjective ratings of sleepiness during a night awake. Sleep. 1994;17(3):236-241.\\

${[23]}$ Tokui S, Oono K, Hido S, Clayton J. Chainer: a next-generation open source framework for deep learning, in Proc. Workshop on Machine Learning Systems in the Annual Conference on Neural Information Processing Systems (NIPS), 2015, 1-6.\\

${[24]}$ Kingma DP, Ba J. Adam: A method for stochastic optimization, in Proc. International Conference on Learning Representations, 2015.\\

${[25]}$ {\"O}zdenizci O, Wang Y, Koike-Akino T, Erdo\u{g}mu\c{s} D. Transfer learning in brain-computer interfaces with adversarial variational autoencoders, in Proc. IEEE EMBS Conference on Neural Engineering, 2019.\\

${[26]}$ {\"O}zdenizci O, Wang Y, Koike-Akino T, Erdo\u{g}mu\c{s} D. Adversarial deep learning in EEG biometrics. IEEE Signal Processing Letters. 2019;26(5):710-714.\\

\end{document}